# The Power of Islamic Scholars' Lecture to Decide Using Islamic Bank with Customer Response Strength Approach

Suryo Budi Santoso*  
Universitas Muhammadiyah Purwokerto  
suryobs@gmail.com

Herni Justiana Astuti  
Universitas Muhammadiyah Purwokerto  
herni99@gmail.com

*Abstract.* **The purpose of this study is to analyze the power of Islamic scholars' lectures to decide using Islamic banks with a customer response strength approach. The sampling technique uses purposive sampling, and the respondents were those who attended lectures on Islamic banks delivered by Islamic scholars. The number of respondents who met the requirements was 96 respondents. Data were analyzed using the customer response strength method. The instrument has met the valid and reliable criteria. The results showed 99% of the total number of respondents acted according to their perceptions of the contents of Islamic banks lectures. Lecture material delivered by scholars about Islamic banks has a strong relationship with their responses ranging from giving attention, interest, fostering desires, and beliefs to having an interest in making transactions with Islamic banks.**

## Introduction

The socialization of Islamic banks has been started since the existence of Islamic banks in Indonesia in 1992. However, the development of Islamic banks in Indonesia is relatively slow when compared to the development of Islamic banks in Malaysia [1]. Therefore, a good strategy is needed in the socialization of Islamic banks in Indonesia [2], [3]. The government itself is aware of this, so the Financial Services Authority (OJK) created a road map for Islamic banking in Indonesia 2015-2019 which contained in number 6 mentioned the need for good and integrated sharia banking socialization with various stakeholders [4], [5].

Based on previous researches, the success of the socialization was influenced by the synergy and Proactive variables [6],[7]. Between the two synergy variables have the greatest effect with a correlation coefficient of 0.464. The synergy variable consists of four influencing variables, namely Business Institutions, Formal Education, Higher Education, and Ulama. Ulama variable influences Synergy, which has the biggest contribution and is equal to 0.290. The contribution is influenced by the lecture material delivered to his followers (worshipers).

Reuters in 2018 noted that of the total Islamic financial assets of the world, 71 per cent were assets of Islamic banks [8]. The condition of Islamic bank assets in Indonesia is a paradox. The country of Indonesia is famous for the most significant number of Muslims in the world [9]. However, the total assets of Islamic banks compared to conventional banks is very minimal, only less than 6 per cent [10]; [11]. Ismal said that Islamic banking in Indonesia still stands on a single-digit market share [12], [13]. The majority of Indonesian Muslims hope that the development of sharia banking in Indonesia will not be like this [14]. Indonesian Islamic bank assets do have growth or development but are still slow when compared to conventional bank assets [6]. Therefore, rules are needed regarding the socialization strategy of Islamic banks in Indonesia so that the development of Islamic bank assets is even faster [6], [15], [16].

Messages delivered through Islamic bank lecture material by the Ulama will be captured by the recipient, namely the congregation/follower. When the message is received, the recipient will respond to the message delivered. The more information obtained, the more likely a person is to form a response. But the recipient does not always look for all the information, usually only pay attention to information that is liked or understood. The appeal of the message conveyed has an impact on consumer responses, and interesting lecture material is certainly easy to get attention. The frequency with which material is given influences audience response. The credibility of lecturers can also affect the credibility of the content of lecture material [17].

## Method

The sampling technique uses purposive sampling, and the respondents were those who attended lectures on Islamic banks delivered by Islamic scholars. The number of respondents who met the requirements was 96 respondents. Data were analyzed using the customer response strength method.

Measurement of the power of lectures uses several procedures:
1) such as finding out the description of the power of the lecture, a calculation is done by multiplying the attractiveness of the lecture, and the frequency of worshipers listening to lecture material on Islamic banks.
2) To determine the consumer response to lecture material used the customer response strength (CRS) method.

In CRS analysis, individual analysis is done by multiplying the length and width dimensions of pilgrim responses. It also sought a customer response index (CRI) / pilgrim response index obtained from the maximum number of stages of the response passed by respondents.

The response scheme used is the AIDCA (attention, interest, desire, conviction dan action) scheme. The formula used is the lecture power formula as described above. Besides, the customer response index (CRI) will also be sought from the maximum number of responses stages the respondent will pass through. All respondents are grouped into those who are exposed (aware) and not exposed then the percentage is made. The process is continued by dividing awareness into what interests and how many do not interest, and so on.

3) To see the relationship between the power of the lecture with the response of the congregation used Spearman rank correlation. In this study, the authors used a 5% significance level. The strength of the response is how much the consumer's reaction to certain stimulants. To measure the strength of the response, we pay attention to the dimensions of the length and width of the response after multiplying the two dimensions, and then the response strength is obtained [17], the formula is as follows:

$$Br = \sum_{i=1}^{n} \Pr_i x L r_i$$

Notes: Br = strength of response; Pri = length of response; N = State the number of stages passed;
Lri = width of the response.

**RESULT & DISCUSSION**

The instrument has met the valid and reliable criteria. All statement items are valid and reliable.

a. Response Index
Table 1 shows the maximum stages that pilgrims go through in processing information from listening to the Ulama's lecture on Islamic banks. Based on table 1, it can be seen that 99% of all pilgrims go through the maximum stages of attention to action from their perceptions of the contents of lectures about Islamic banks.

b. The Power of Lecture
To determine the strength of lectures that can affect information processing in worshipers, the average value of respondents' perceptions of the strength of lectures can be calculated, which can be seen as shown in Table 2.

Table 1. Index of Respondents' Responses to the Power of Lecture

| Maximum Stage that was passed | Power of Response | Frequency | Total (%) |
|---|---|---|---|
| Attention | NAtt | 1 | 1 |
| | Att1 | 14 | 99 |
| | Att2 | 53 | |
| | Att3 | 28 | |
| Interest | NInt | 1 | 1 |
| | Int1 | 27 | 99 |
| | Int2 | 48 | |
| | Int3 | 20 | |
| Desire | NDes | 1 | 1 |
| | Des1 | 18 | 99 |
| | Des2 | 58 | |
| | Des3 | 19 | |
| Conviction | NCon | 1 | 1 |
| | Con1 | 10 | 99 |
| | Con2 | 46 | |
| | Con3 | 39 | |
| Action | NAct | 1 | 1 |
| | Act1 | 23 | 99 |
| | Act2 | 53 | |
| | Act3 | 19 | |

Table.2 Average Value of Lecture Strength

| | The appeal of lectures | Response pilgrims | Total |
|---|---|---|---|
| Sum | 1237 | 1943 | 3180 |
| Mean | 5,43 | 4,05 | 4,14 |

Table 2 shows the attractiveness of the lecture content and how much the pilgrims responded. Based on Table 2, it can be seen that the average value of the pilgrim's response is 4.05, the average value of lecture attraction is 5.43, so the average value of lecture strength is 4.14. The average value of the lecture's strength is in the interval of values between 1.61 to 4.80. Thus, the magnitude of the average strength of lectures is in a strong category.

c. Power of Response
Analyzing the strength of responses to lecture is done by finding the average value of respondents' perceptions of response strength, namely with the formula CRS (Score Attention x 1) + (Interest Score x 2) + (Score Desire x 3) + (Conviction Score x 4) + (Action Score x 5) which can be seen in Table 3.

Table 3. Average Customer Response Strength

| | Attention | Interest | Desire | Conviction | Action | Sum |
|---|---|---|---|---|---|---|
| Sum | 396 | 375 | 383 | 411 | 378 | 1943 |
| Mean | 4,125 | 3,906 | 3,989 | 4,281 | 3,937 | 20,24 |

Based on Table 3, it can be seen that the average value of the respondent's attention at the lecture is 4.125, the average value of the respondent's interest in the lecture is 3.906, the average desire of the respondent at the lecture is 3.989, the average conviction of the respondent at the lecture is 4.281 and the average respondent's action on the lecture was 3,937 so that the average value of the response strength on the lecture was 20.24. The average strength of the response is at an interval of less than 27.01. Thus, the magnitude of the average strength of the response of pilgrims to lectures based on the method of Customer Response Strength is in the category of strong enough.

d. The relationship between the power of the lecture and the response of the pilgrims

Based on the calculation shows that the value of r count Spearman rank correlation is 0.614, with a significance value of 0.000 smaller than α (0.05). The r count value is in the interval of values between 0.60 to 0.79. It shows that there is a strong positive relationship between the power of lectures on Islamic banks with the strength of the response of pilgrims.

The recipient, the pilgrims, will capture messages delivered through Islamic bank lecture material by the Ulama. When the message is received, the recipient will respond to the message delivered. The more information obtained, the more likely a person is to form a response. But the recipient does not always look for all information, usually only pay attention to information that is liked or understood [17]. The attractiveness of the message conveyed has an impact on consumer responses, and interesting lecture material is certainly easy to get attention [17].

In general, pilgrims as respondents in this study gave a good response to lectures on Islamic banks delivered by scholars. The good response is shown by the high level of attention (attention) of the respondents who then foster interest (interest), where the strong interest then fosters hopes (desire) towards Islamic banks. Information about Islamic banks will further encourage confidence (conviction) and be realized through action, both in the form of interest in making transactions and making transactions. The power of lectures delivered by Ulama also has a strong relationship with the response of pilgrims.

## CONCLUSION

Lecture material delivered by scholars to their congregation about Islamic banks has a strong relationship with their responses ranging from giving attention, interest, fostering desires and beliefs to having an interest in making transactions with Islamic banks. This was also strengthened by using the customer response strength method.